\documentclass[varenna]{cimento}

\usepackage{bm}
\usepackage[dvips]{graphicx}
\usepackage{amssymb}

\def\be{\begin{equation}}
\def\ee{\end{equation}}

\def\ua{\uparrow}
\def\da{\downarrow}

\title{Unitary polarized Fermi gases}

\author{F. Chevy}

\institute{\'Ecole normale sup\'erieure, 24, rue Lhomond, Paris,
France}

\begin{document}

\maketitle

\begin{abstract}
Although recent theoretical and experimental progress have
considerably clarified pairing mechanisms in spin 1/2 fermionic
superfluid with equally populated internal states, many open
questions remain when the two spin populations are mismatched. We
show here that, taking advantage of the universal behavior
characterizing the regime of infinite scattering length, the
macroscopic properties of these systems can be simply and
quantitatively understood in the regime of strong interactions.
\end{abstract}

\section{Introduction}

Pairing lies at the core of the standard Bardeen-Cooper-Schrieffer
mechanism for metal superconductivity, and the very natural
question to know whether it could survive population imbalances
between the two spin states naturally arose very soon after its
development \cite{Clogston62,Chandrasekhar62}. It was pointed out
that pairing was indeed robust to some amount of mismatch between
the chemical potentials of the two species, but the fate of the
system after the critical imbalance is reached has long been a
mystery. The absence of clear answer to this problem was due in
particular to the absence of an experimental system on which the
various scenarios envisioned could be tested: existence of a
spatially modulated order parameter (Fulde, Ferrel, Larkin and
Ovshinikov, or FFLO, phases) \cite{Sarma63,Fulde64,Larkin65}, or
the extension to trapped
systems\cite{Combescot05,Mora05,Castorina05,Mizushima05}, deformed
Fermi surfaces \cite{Sedrakian05}, interior gap superfluidity
\cite{Liu03}, phase separation between a normal and a superfluid
state through a first order phase transition
\cite{Bedaque03,Caldas05,Carlson05,Cohen05}, BCS quasi-particle
interactions \cite{Ho06} or onset of p-wave pairing
\cite{Bulgac06b}. When the strength of the interactions is varied,
a complicated phase diagram mixing several of these scenarios is
expected \cite{Pao05,Son05,Sheehy05}.

This issue was revived by the possibility of obtaining fermionic
superfluids with ultra cold atoms
\cite{Jochim03,Zwierlein03,Greiner03,Bourdel04,Kinast04,Partridge05b},
where spin imbalance could be controlled and maintained for a long
time. This led to a series of experiments performed at MIT
\cite{Zwierlein05,Zwierlein06} and Rice
\cite{Partridge05,Partridge06c} which clearly demonstrated a phase
separation between regions characterized by different
polarizations (i.e. spin population imbalances, by analogy with
magnetism). The number of phases obtained by the two groups is
however different. In Rice experiment, the cloud is constituted of
a core where both spin populations are equal, surrounded by a
shell of majority atoms only while at MIT a third phase mixing
both species with unequal densities is intercalated between the
previous ones, a discrepancy which is not yet fully explained
\cite{Silva06,Imambekov06,Imambekov06,Pieri05,Yi06,Haque06,Bulgac06,Chevy06,Chevy06b}.

In what follows we wish to explore the various consequences of
these experiments. By contrast to most recent works on the
subject, we would like to avoid the use of BCS mean field, which
is known to give good qualitative insight to the problem under
study, but fails when precise quantitative estimates are needed.
Our scheme is based on a combination of exact variational analysis
and Monte-carlo simulations. We will demonstrate that, in
agreement with MIT experiments, three phases are expected in
homogeneous systems. To compare with experimental results, we will
make use of Local Density Approximation (LDA) which leads to
quantitative agreement with MIT's data. Finally, following
\cite{Silva06}, we will show how Rice's apparently contradictory
results can be interpreted as a breakdown of local density
approximation in elongated traps.

\section{Universal phase diagram of a homogeneous system}

Let us first consider an ensemble of spin 1/2 fermions of mass $m$
trapped in a box of volume $V$. In the s-wave approximation, the
hamiltonian $\widehat H$ is given by

\be \widehat H=\sum_{\bm k,\sigma}\epsilon_{\bm k}\widehat
a^\dagger_{\bm k,\sigma} \widehat a_{\bm
k,\sigma}+\frac{g_b}{V}\sum_{\bm k,\bm k',\bm q}\widehat
a^\dagger_{\bm k+\bm q,\uparrow}a^\dagger_{\bm k'-\bm q
,\downarrow} \widehat a_{\bm k',\downarrow}\widehat a_{\bm
k,\uparrow}, \ee

\noindent where $\epsilon_{\bm k}=\hbar^2k^2/2m$, $\widehat a_{\bm
k,\sigma}$ annihilates a particle of spin $\sigma$ and momentum
$\bm k$ and $g_b$ is the coupling constant characterizing s-wave
interactions between atoms. This choice of interaction potential
is singular and yields unphysical results and to get rid of the
divergencies resulting by the zero range of the potential, we
introduce an ultraviolet cut-off $q_c$ in momentum space (or
equivalently, we work on a lattice of step $1/q_c$). When $q_c$
goes to infinity, the Lippmann-Schwinger formula obtained by the
resolution of the two-body problem yields the following
relationship between the bare coupling constant and the scattering
length $a$

\be \frac{1}{g_b}=\frac{m}{4\pi\hbar^2a}-\frac{1}{V}\sum_{\bm
k}\frac{1}{\epsilon_{\bm k}}, \label{Eqn7}\ee

\noindent where the sum over $\bm k$ is restricted to $k<q_c$.

To anticipate the analysis of inhomogeneous systems, we work in
the grand canonical ensemble, where the atom numbers fluctuate and
only their expectation values are kept constant. Introducing the
chemical potentials $\mu_{\uparrow,\downarrow}$ as Lagrange
multipliers associated with the constraints on atom numbers, we
 need to find the ground state of the grand potential
$\widehat \Omega$ given by

\be \widehat\Omega=\widehat H-\mu_\ua\widehat
N_\ua-\mu_\da\widehat N_\da. \ee

In what follows, we replace the minimization condition on
$\Omega=\langle\widehat\Omega\rangle$ by a maximization problem on
the pressure $P$, using the thermodynamical relation $\Omega=-PV$.
Moreover, we assume $\mu_\ua>\mu_\da$ and we restrict ourselves to
the unitary limit where $a=\infty$. This choice of scattering
length leads to a deep simplification of the formalism, due to the
universality characterizing this regime. Indeed, from dimensional
analysis \cite{Vaschy92}, we can show that for an arbitrary
scattering length, the pressure $P$ of a given phase is given by
some relation

$$P(m,\hbar,a,\mu_\ua,\mu_\da)=P_0(\mu_\ua,\hbar,m) f(\mu_\da/\mu_\ua,1/k_{\rm
F\ua}a),$$

\noindent where $P_0$ is the pressure of an ideal Fermi gas with
chemical potential $\mu_\ua$ and $k_{F\ua}$ is the Fermi wave
vector associated with $\mu_\ua$. At unitarity, $1/k_{\rm F}a=0$
and $f$ is therefore function of $\eta=\mu_\da/\mu_\ua$ yielding
the universal relation

\be \frac{P}{P_0}=g(\eta), \ee

\noindent where $g(\mu_\da/\mu_\ua)=f(\mu_\da/\mu_\ua,0).$

Although the general minimization of the grand potential is an
extremely challenging and still open problem, we first note that
two exact eigenstates of the system can be found.

\begin{enumerate}
\item {\em Fully polarized ideal gas}. If we consider a fully
polarized system containing no minority atom, the interaction term
in $\widehat H$ disappears, and we are left with a pure ideal gas
of majority atoms. The pressure of this normal phase is simply the
Fermi pressure, and we have in particular $P/P_0=1$.

\item {\em Fully paired superfluid}. Let $|{\rm SF}\rangle_\mu$ be
the ground state of the {\em balanced} potential $\widehat
\Omega'=\widehat H-\mu(\widehat N_\ua+\widehat N_\da)$. Since
$\widehat\Omega'$ commutes with the atom number operators,
 $|{\rm SF}\rangle_\mu$ can be chosen as an eigenstate of both $\widehat
 N_{\ua,\da}$, with $\widehat N_\ua |{\rm SF}\rangle_\mu=\widehat N_\da |{\rm
 SF}\rangle_\mu$. Going back to the unbalanced problem, we write $\widehat\Omega$ as

 \be\widehat\Omega=\widehat H+\frac{\mu_\ua+\mu_\da}{2}(\widehat N_\ua+\widehat
 N_\da)+\frac{\mu_\ua-\mu_\da}{2}(\widehat N_\ua-\widehat
 N_\da).\ee

We see readily that for $\mu=(\mu_1+\mu_2)/2$ we have $\widehat
\Omega|{\rm
 SF}\rangle_{\mu}=\widehat\Omega'|{\rm SF}\rangle_\mu$, which
 proves that $|{\rm SF}\rangle_\mu$ is also an eigenstate of the
 imbalanced grand potential. The pressure in this superfluid phase can be calculated using known results
 for the unitary balanced superfluid for which the universal relationship between chemical potential and density
 reads

 \be
 \mu_\ua=\mu_\da=\xi\frac{\hbar^2}{2m}(6\pi^2 n_\ua)^{2/3},
 \ee

\noindent where $\xi\sim 0.42$ is a universal number that was
evaluated both experimentally
\cite{Bourdel04,Bartenstein04,OHara02,Kinast05,Partridge05}
 and
theoretically \cite{Carlson03,Perali04,Astrakharchik04,Carlson05}.
Integrating Gibbs-Duhem identity (see appendix), one then obtains
 for the imbalanced system

 \be P_{\rm
 SF}=\frac{1}{15\pi^2}\left(\frac{m}{\xi\hbar^2}\right)^{3/2}(\mu_\ua+\mu_\da)^{5/2},\ee

 \noindent hence $P_{\rm SF}/P_0=(1+\eta)^{5/2}/(2\xi)^{3/2}.$
 \end{enumerate}

\begin{figure}
\centerline{\includegraphics{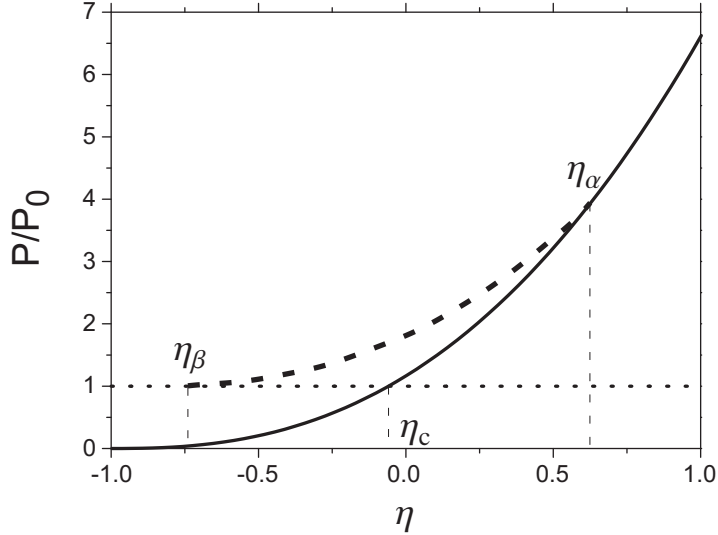}} \caption{Sketch of the
grand potential $\Omega$ as a function of $\eta=\mu_\da/\mu_\ua$.
$\Omega$ is normalized to the grand potential $\Omega_0$ of the
pure ideal gas of chemical potential $\mu_\ua$. Full line: paired
superfluid; dotted line: fully polarized normal phase; dashed
line: intermediate mixed phase. $\eta_\alpha$ and $\eta_\beta$
designate the critical values for the two transitions between the
superfluid/mixed phase and the mixed phase/fully polarized Fermi
gas.} \label{Fig1}
\end{figure}

The variation of the pressure versus $\eta$ is displayed in Fig.
\ref{Fig1}. We see that for small imbalances, {\em i.e.} $\eta$
smaller than $\eta_c=(2\xi)^{3/5}-1\sim -0.10$, the fully paired
superfluid is more stable than the fully paired normal phase,
confirming the stability of pairing against a small mismatch of
the Fermi surfaces. The experimental results presented in ref.
\cite{Zwierlein06} suggest that the two classes of states we have
until now restricted ourselves are not sufficient to fully capture
the physics of imbalanced systems. In particular, a mixed normal
phase, containing atoms of both species in unequal proportions,
must be taken into account. A sketch of $g(\eta)$ for this
intermediate phase is shown in Fig \ref{Fig1}. On this more
general phase diagram, the parameters $\eta_\alpha$ and
$\eta_\beta$ are of special importance, since they characterize
the phase transitions between the three different phases. A glance
at Fig. \ref{Fig1} shows that they must satisfy the inequality

$$\eta_\alpha<\eta_c<\eta_\beta,$$

\noindent and the next section is devoted to an improvement of
these bounds.

\section{The N+1 body problem}

Theoretically, the existence of the intermediate phase can be
demonstrated by the study of the N+1 body problem, in other word
the study of the ground state of the majority Fermi sea in the
presence of a single minority atom. This particular system
corresponds to an intermediate phase with
$\eta\rightarrow\eta_\beta^+$ and we will prove that it yields the
inequality $\eta_\beta<\eta_c$.

 To address the N+1 body problem, we use a variational scheme, that we will compare to recent predictions based on Monte-Carlo simulations \cite{Lobo06}. Let us consider the following trial
state $|\psi\rangle$

$$|\psi\rangle=\phi_0|{\rm FS}\rangle+\sum_{\bm k,\bm q}\phi_{\bm k,\bm
q}|\bm k,\bm q\rangle,$$

\noindent where $|{\rm FS}\rangle$ is a spin up Fermi sea plus a
spin down impurity with 0 momentum, and $|\bm k,\bm q\rangle$ is
the perturbed Fermi sea with a spin up atom with momentum $\bm q$
(with $q$ lower than $k_F$) excited to momentum $\bm k$ (with
$k>k_F$). To satisfy momentum conservation, the impurity acquires
a momentum $\bm q-\bm k$.

The energy of this state with respect to the non interacting
ground state is $\langle \widehat H\rangle=\langle\widehat
H_0\rangle+\langle\widehat V\rangle$, with

$$\langle\psi|\widehat H_0|\psi\rangle=\sum_{\bm k,\bm q}|\phi_{\bm k,\bm
q}|^2(\epsilon_{\bm k}+\epsilon_{\bm q-\bm k}-\epsilon_{\bm q}),$$

\noindent and

$$\langle\psi|\widehat V|\psi\rangle=\frac{g_B}{V}\left(\sum_{\bm
q}|\phi_0|^2+\sum_{\bm k,\bm k',\bm q}\phi_{\bm k',\bm
q}\phi^*_{\bm k,\bm q}+\sum_{\bm k,\bm q,\bm q'}\phi_{\bm k,\bm
q}\phi^*_{\bm k,\bm q'}+\sum_{\bm q,\bm k}(\phi^*_0\phi_{\bm k,\bm
q}+\phi_0\phi^*_{\bm k,\bm q})\right),$$

\noindent where $\epsilon_{\bm k}=\hbar^2 k^2/2$, and the sums on
$q$ and $k$ are implicitly limited to $q<k_F$ and $k>k_F$. As we
will check later, $\phi_{\bm k,\bm q}\sim 1/k^2$ for large momenta
(see below, eqn. (\ref{Eqn3})), in order to satisfy the short
range behavior $1/r$ of the pair wave function in real space. This
means that most of the sums on $\bm k$ diverge for
$k\rightarrow\infty$. This singular behavior is regularized by the
renormalization of the coupling constant $g_B$ using  the
Lippman-Schwinger formula. It implies that $g_B$ vanishes for
large cutoff, thus yielding a finite energy. However, it must be
noted that the third sum in $\langle\psi|\widehat V|\psi\rangle$
is convergent and when multiplied by $g_B$ will give a zero
contribution to the final energy and can therefore be omitted in
the rest of the calculation.

The minimization of $\langle \widehat H\rangle$ with respect to
$\phi_0$ and $\phi_{\bm k,\bm q}$ is straightforward and yields
the following set of equations

\begin{eqnarray}
\frac{g_B}{V}\sum_{\bm q}\phi_0+\frac{g_B}{V}\sum_{\bm q,\bm
k}\phi_{\bm k,\bm q}&=&E\phi_0\label{Eqn1}{}\\
(\epsilon_{\bm k}+\epsilon_{\bm q-\bm k}-\epsilon_{\bm
q})\phi_{\bm k,\bm q}+\frac{g_B}{V}\sum_{\bm k'}\phi_{\bm k',\bm
q}+\frac{g_B}{V}\phi_0&=&E\phi_{\bm k,\bm q},\label{Eqn2}
\end{eqnarray}

\noindent where $E$ is the Lagrange multiplier associated to the
normalization of $|\psi\rangle$, and can also be identified with
the trial energy. Let us introduce $\chi(\bm q)=\phi_0+\sum_{\bm
k}\phi_{\bm k,\bm q}$. We see from eqn. \ref{Eqn2} that

\begin{equation}
\phi_{\bm k,\bm q}=-\frac{g_B}{V}\frac{\chi({\bm
q})}{\epsilon_{\bm k}+\epsilon_{\bm q-\bm k}-\epsilon_{\bm
q}-E}.\label{Eqn3}\end{equation}

As expected, we note here the $1/\epsilon_{\bm k}\sim 1/k^2$
dependence for large $k$. Inserting this expression in the
definition of $\chi$, we obtain

$$\chi (\bm q)=\phi_0-\frac{g_B}{V}\sum_{\bm
k}\frac{\chi(\bm q)}{\epsilon_{\bm k}+\epsilon_{\bm q-\bm
k}-\epsilon_{\bm q}-E},$$

\noindent that is

$$\chi(\bm q)=\frac{\phi_0/g_B}{\frac{1}{g_B}+\frac{1}{V}\sum_{k>k_F}\frac{1}{\epsilon_{\bm k}+\epsilon_{\bm q-\bm
k}-\epsilon_{\bm q}-E}}$$

Finally, eqn. (\ref{Eqn1}) can be recast as
$E\phi_0/g_B=\sum_{q<k_F}\chi(\bm q)/V$, that is, using the
explicit expression for $\chi (\bm q)$:

$$E=\frac{1}{V}\sum_{q<k_F}\frac{1}{\frac{1}{g_B}+\frac{1}{V}\sum_{k>k_F}\frac{1}{\epsilon_{\bm k}+\epsilon_{\bm q-\bm
k}-\epsilon_{\bm q}-E}}.$$

We get rid of the bare coupling constant $g_B$ by using the
Lippman-Schwinger equation, which finally yields the following
implicit equation for $E$

\begin{equation}
E=\frac{1}{V}\sum_{q<k_F}\frac{1}{\frac{m}{4\pi\hbar^2a}-\frac{1}{V}\sum_{k<k_F}\frac{1}{2\epsilon_{\bm
k}}+\frac{1}{V}\sum_{k>k_F}\left(\frac{1}{\epsilon_{\bm
k}+\epsilon_{\bm q-\bm k}-\epsilon_{\bm
q}-E}-\frac{1}{2\epsilon_{\bm k}}\right)}. \label{Eqn4}
\end{equation}

Before addressing the unitary limit case, let us show that this
formula allows us to recover the known exact results in the limit
of small scattering lengths where the denominator is dominated by
the $1/a$ term. The correction to the energy is therefore

$$E\sim \frac{1}{V}\sum_{q<k_{F}}\frac{4\pi\hbar^2a}{m}=\frac{4\pi\hbar^2a}{m}\frac{N}{V}$$

\noindent where $N$ is the total number of majority atoms. We thus
see that the trial state recovers the mean-field prediction for
low interactions. For $a\rightarrow 0^+$ (BEC regime), a little
more involved calculation allows one to recover the classical
molecular binding energy $E\sim -\hbar^2/ma^2$. Finally in the
case of the unitary regime relevant to experiments, eqn.
(\ref{Eqn4}) is solved numerically and yields $E\sim -0.3\hbar^2
k_F^2/m$, that is $\eta_\beta<-0.60$, a value remarkably close to
that obtained in Monte-Carlo simulations \cite{Lobo06}.

\section{Trapped system and comparison with MIT experiment}

The model presented in the previous section adresses only the
situation of a homogeneous system and to compare with experiments,
we need to extend the formalism developed in the previous section
to the case of trapped systems. To this purpose we make use of the
Local Density Approximation (LDA), in which we assume that the
chemical potential of species $\sigma$ varies as

\be \mu_\sigma(\bm r)=\mu_\sigma^0-V(\bm r), \ee

\noindent where $V$ is the trapping potential. From this relation,
we see that varying $r$ is equivalent to varying the chemical
potentials of the two species, and in particular their ratio
$\eta(r)$ The two phase transitions described in the previous
section will happen for radii $r=R_{\alpha,\beta}$ such that
$\mu_\da({R_{\alpha,\beta}})/\mu_\ua({R_{\alpha,\beta}})=\eta_{\alpha,\beta}$.
Moreover, since the outer rim is constituted by a normal ideal
gas, the boundary $R_\ua$ of the majority component is given by
the condition $\mu_\ua(R_\ua)=0$.

In an isotropic harmonic trap, we can combine these three
relations to eliminate the parameters $\mu_\sigma^{(0)}$, thus
obtaining the general relation relating the three radii
$R_{\alpha,\beta,\ua}$:

\be
\left(\frac{R_\alpha}{R_\ua}\right)^2=\frac{(R_\beta/R_\ua)^2-q}{1-q},
\label{Eqn9}\ee

\noindent where $q=(\eta_\alpha-\eta_\beta)/(1-\eta_\beta)$. One
striking consequence of this equation is the prediction of a
threshold at which $R_\alpha$ vanishes, corresponding to the
disappearance of the fully paired superfluid. This transition
happens when the ratio $(R_\beta/R_\ua)^2$ reaches the critical
value $q$. From the upper and lower bounds obtained for
$\eta_\alpha$ and $\eta_\beta$, we see that $q>0.30$.

\begin{figure}
\centerline{\includegraphics{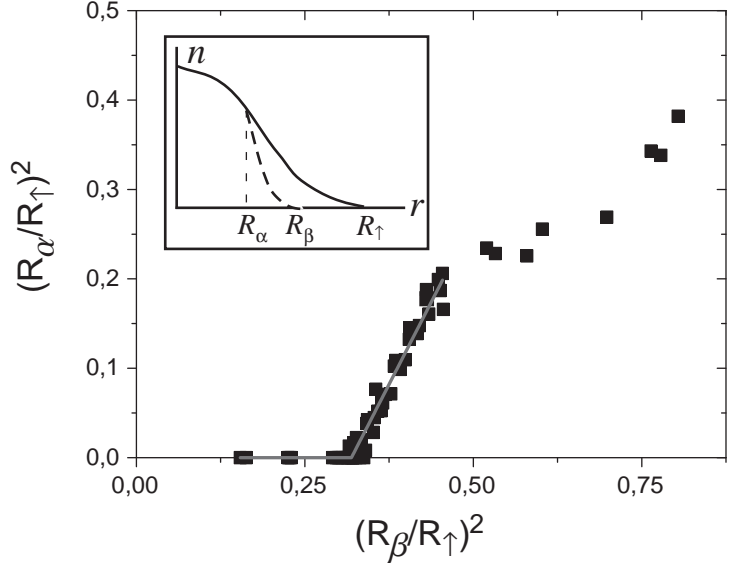}} \caption{Comparison of
equation (\ref{Eqn9}) with experimental data from MIT.  A fit to
the data yields $q\sim 0.32$. Inset: sketch of the density
profile. The full (resp. dashed) line corresponds to the density
of the majority (resp. minority) component. $R_\alpha$ marks the
end of the superfluid region, $R_\beta$ that of the mixture and
$R_\uparrow$ is the frontier of the majority cloud.} \label{Fig2}
\end{figure}

This prediction of LDA is remarkably well verified in MIT's
experiments \cite{Zwierlein06} for which the three phases
discussed above were indeed observed, and  eq. (\ref{Eqn9}) could
be tested experimentally (Fig. \ref{Fig2}). On this graph, we see
that for large imbalance, the linear scaling predicted by eq.
(\ref{Eqn9}) is indeed satisfied, with $q\sim 0.32$, in agreement
with the lower bound obtained earlier. The deviation from theory
observed for $(R_\beta/R_\ua)^2\gtrsim 0.5$ is not yet fully
understood. However, it must be noted that the discrepancy takes
place in a regime of low imbalance, where the phase transitions
take place in the tail of the density distribution. In these
regions of low density, we may observe a breakdown of the LDA, or
of the hydrodynamical expansion that was used to infer the
experimental radii.

The value $q\sim 0.32$ obtained from the comparison with
experimental data can help us improve the bounds for
$\eta_{\alpha,\beta}$. Indeed, this relation fixes the relative
values of $\eta_\alpha$ and $\eta_\beta$. When combined with the
bounds found in the previous section,  we obtain indeed

\begin{eqnarray}
\label{Eqn10} -0.62&<\eta_\beta<&-0.60\\
\label{Eqn11}-0.10&<\eta_\alpha<&-0.088
\end{eqnarray}

From the previous analysis, we see that the combination of
theoretical arguments and analysis of experimental data allows for
a precise determination of the thresholds of the different phase
transitions. Knowing the values of $\eta_{\alpha,\beta}$ as well
as the exact equation of state in the fully polarized and fully
paired phases, we can even obtain some upper and lower bounds for
the equation of state of the mixed phase, using the concavity of
the grand potential \cite{Bulgac06}.

\section{Elongated systems and Rice's experiment}

\begin{figure}
\centerline{\includegraphics{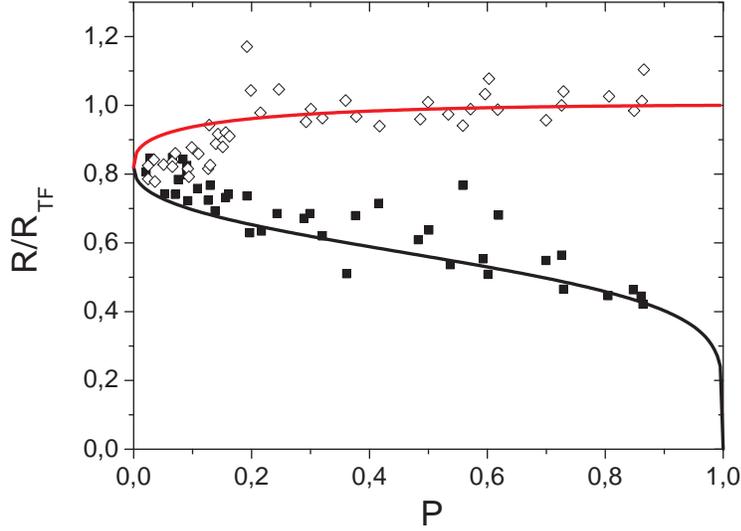}} \caption{Rice's radius
measurement and comparison with a two phase model. The radius $R_i$
is scaled in units of the Thomas Fermi radius of an ideal gas with a
the same atom number $N_i$.} \label{Fig3}
\end{figure}

Surprisingly, similar experiments performed at Rice University
showed no evidence of an intermediate phase, but rather the
coexistence of the fully paired and fully polarized phases only.
Measurements of the axial radii of the two phases from ref.
\cite{Partridge05} are presented in Fig. \ref{Fig3} and can be
compared with the model presented above when omitting the
intermediate mixed phase \cite{Chevy06}. In these conditions, the
inner superfluid region is now defined by the condition $\mu_\da(\bm
r)/\mu_\ua (\bm r)<\eta_c$ and is bounded by the radius $R_\da$
defined by

\be R_\da^2=\frac{2}{m\bar\omega^2}\left(\frac{\mu_\da^0-\eta_{\rm c
}\mu_\ua^0}{1-\eta_{\rm c}}\right). \ee

Atoms of the minority species are located in the paired superfluid
phase only. We thus have

\be N_\da=\int_{r<R_\da}n_\da (\bm r)\,d^3\bm
r=\frac{2}{3\pi\xi^{3/2}}\left(\frac{\mu_\ua^0+\mu_\da^0}{\hbar\bar\omega}\right)^3g(R_\da/\bar
R), \label{Eqn4a}\ee

\noindent where $\bar R^2=(\mu_\ua^0+\mu_\da^0)/m\bar\omega^2$ and

\be g(x)=\frac{x\,{\sqrt{1 - x^2}}\,
     \left( -3 + 14\,x^2 - 8\,x^4 \right)  + 3\,\arcsin (x)}
    {48}.
\ee

Excess atoms of the majority species are located between $r=R_\da$
and $r=R_\ua$ such that $m\bar\omega^2 R_\ua^2/2=\mu_\ua^0$. The
number of excess atoms is therefore
$N_\ua-N_\da=\int_{R_\da}^{R_\ua} n_1 (\bm r)\, d^3\bm r$, hence

\be
N_\ua-N_\da=\frac{2}{3\pi}\left(\frac{2\mu_\ua^0}{\hbar\bar\omega}\right)^3(g(1)-g(R_\da/R_\ua)).
\label{Eqn5a}\ee

Dividing by (\ref{Eqn5a}) by (\ref{Eqn4a}) yields the implicit
equation for $\eta_0=\mu^0_\da/\mu_\ua^0$ as a function of
$N_\ua/N_\da$

\be
\frac{N_\ua}{N_\da}=1+\xi^{3/2}\frac{8}{(1+\eta_0)^3}\frac{g(1)-g(R_\da/R_\ua)}{g(R_\da/\bar
R)}. \label{Eqn6a}\ee

Equation (\ref{Eqn6a}) is  solved numerically and the value obtained
for $\eta_0$ is then used to calculate the radii $R_\ua$ and
$R_\da$. The predicted evolution of the $R_i$ versus the population
imbalance $P=(N_\ua-N_\da)/(N_\ua+N_\da)$ is shown in Fig.
\ref{Fig2}. To follow Ref. \cite{Partridge05}, we have normalized
each $R_i$ to the Thomas-Fermi radius $R_{\rm TF}$ associated to an
ideal gas containing $N_i$ atoms. The agreement with the
experimental data is quite good as soon as $P\gtrsim 0.1$, a
remarkable result, since the model presented here contains no
adjustable parameter, as soon as the value of $\xi$ is known.

Despite this remarkable agreement, this simple two phase+local
density approximation model fails to captures all experimental
features. In particular, a qualitative discrepancy occurs in the
comparison between the theoretical and integrated density
profiles. Indeed, as shown in \cite{Silva06a}, LDA at unitarity
implies a constant density difference in the paired superfluid
region, in contradiction with experimental data. One solution to
this problem was presented in \cite{Imambekov06,Silva06}. In these
papers, it is noted that in the presence of phase transitions, the
description of the sharp frontier separating to adjacent phases
involves the introduction of density gradient terms in the energy.
When the interface in thin enough, they can be encapsulated in a
new surface tension energy term reading \cite{Silva06}

$$\Omega_{\rm ST}=\int_{\cal S} \gamma (\mu_{\uparrow\downarrow}(\bm r))d^2S,$$

\noindent where $\cal S$ is the interface between the two phases,
and $\gamma$ is the surface tension constant, which should
dimensionally vary as

$$\gamma=\lambda
\frac{m\mu^2_\uparrow}{\hbar^2}.$$

\begin{figure}
\centerline{\includegraphics{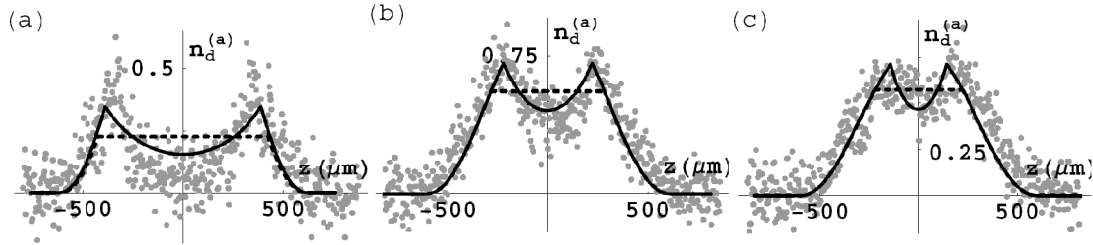}} \caption{Integrated
density difference in Rice experiments, and comparison with the
surface tension model (data from \cite{Silva06}). Dashed line, LDA
prediction: the density difference is flat in the superfluid
region, in contradiction with experimental date. Full line: Two
phase model incorporating surface tension effect. The same surface
tension parameter $\lambda=\sim 10^{-4}$ is used for all three
graphs.} \label{Fig5}
\end{figure}

Here, $\lambda$ is a numerical factor that will be determined by
comparison with experiments and we have used the fact that at the
coincidence between the phases, the ratio
$\mu_\downarrow/\mu_\uparrow$ is fixed and equal to $\eta_c$,
meaning that the two chemical potentials are no longer
independent. We can minimize the total grand potential
$\Omega=\Omega_{\rm bulk}+\Omega_{\rm ST}$, where $\Omega_{\rm
bulk}=-\int\left(P_{\rm N}+P_{\rm SF}\right)d^3r$ is the bulk
contribution to the energy. Following \cite{Silva06} we simplify
the analysis by assuming that the interfaces are ellipsoidal, and
for $\lambda\sim 10^{-4}$ one obtains the results presented in
Fig. \ref{Fig5}, which coincides with experimental data. The
absence of capillary effects at MIT can be explained by a smaller
trap aspect ratio and a larger atom number of atoms compared with
Rice's experimental situation, as shown by a simple scaling
argument \cite{Silva06}.

\section{Conclusion}

The formalism presented here allows for a simple and quantitative
description of {\em macroscopic} properties of  polarized Fermi
gases in the regime of strong interaction. This analysis is
nevertheless far from being complete, since it does not give any
information on the superfluid nature of the various phases. For
instance, the mixed region of the phase diagram may contain
superfluid and  normal subdomains, the transition between this two
regimes being characterized by a universal number
$\eta_\gamma\in[\eta_\beta,\eta_\alpha]$. The quantitative
understanding of these superfluid properties will require beyond
mean-field theories, such as the Monte-Carlo calculations of
\cite{Lobo06}.

\section{Acknowledgments}

The author gratefully acknowledges support by the IFRAF institute
and the ACI Nanosciences 2004 NR 2019. The author thanks the ENS
ultracold atoms group, S.~Stringari, C. Lobo, A. Recati, A.
Bulgac, E.A. Mueller, X. Leyronas, C. Mora and R.~Combescot for
stimulating discussions. Laboratoire Kastler Brossel is a research
unit No. 8552 of CNRS, ENS, and Universit\'e Paris 6.

\section{Appendix: thermodynamical relations for the grand potential}

Let us consider a homogeneous many-body system characterized by a
hamiltonian $\widehat H_0$ and containing particles of $p$
different species labelled by $i=1..p$. In the grand canonical
ensemble, one looks for the ground state of this system by letting
the atom numbers fluctuate, but keeping the expectation values
$\langle\widehat N_{i=1..p}\rangle$ fixed. This therefore requires
to find the ground state of the grand potential
$\widehat\Omega(\mu_i)=\widehat H-\sum_{i=1}^p\mu_i\widehat N_i$,
where the $\mu_i$ are Lagrange multiplier that we interpret as
chemical potentials.

Let $|\psi(\mu_i,V)\rangle$ be the ground state of the grand
potential, we set $\Omega(\mu_i,V)=\langle\psi(\mu_i,V)|\widehat
\Omega|\psi(\mu_i,V)\rangle$. Using Hellman-Feynman relation, we
can write that

\be\frac{\partial\Omega}{\partial\mu_i}=\langle\psi(\mu_i,V)|\frac{\partial\widehat\Omega}{\partial\mu_i}|\psi(\mu_i,V)\rangle=-\langle\widehat
N_i\rangle,\ee

\noindent from which we deduce that

\be d\Omega=\sum_i -N_id\mu_i+\frac{\partial\Omega}{\partial V}dV.
\label{Eqn1b} \ee

By definition, and by analogie with classical thermodynamics, we
identify $\partial_V\Omega$ with $-P$, the pressure in the system.

Let us now us the extensivity of the potential: when the volume is
multiplied by some scaling factor $\lambda$, $\Omega$ is
multiplied by the same factor. In other words, we have $\Omega
(\lambda V,\mu_i)=\lambda\Omega (V,\mu_i)$. Taking $\lambda=1/V$,
we get $\Omega(V,\mu_i)=V\Omega (1,\mu_i)$. Differentiating this
with respect to $V$, we note that $\Omega (1,\mu_i)=-P$, hence

\be \Omega=-PV \label{Eqn2b} \ee

From this equation, we see that the minimum grand potential is
state has also the highest pressure. $P$ can moreover be
calculated by differentiating $\Omega$ and using equations
(\ref{Eqn2b}) and (\ref{Eqn1b}). We then obtain the Gibbs-Duhem
relation

\be dP=\sum_i n_i d\mu_i, \label{Eqn3b} \ee

\noindent where $n_i=N_i/V$ is the density of species $i$. From
equation (\ref{Eqn3b}), we see that the pressure (hence the grand
potential) can be obtained simply from the knowledge of the
equation of state $n_i (\mu_j)$.

\subsection{Concavity}

Since, by definition, $|\psi(\mu_i)\rangle$ is the ground state of
$\widehat \Omega (\mu_i)$, we have for any $\delta\mu_i$

\be \langle\psi (\mu_i+\delta\mu_i)|\widehat\Omega
(\mu_i)|\psi(\mu_i+\delta\mu_i)\rangle \ge \langle\psi
(\mu_i)|\widehat\Omega (\mu_i)|\psi(\mu_i)\rangle \ee

Moreover, if one notes that $\widehat\Omega
(\mu_i)=\widehat\Omega(\mu_i+\delta\mu_i)+\sum_j\delta\mu_j\widehat
N_j$, we see that for any $\delta\mu_i$

\be
\Omega(\mu_i+\delta\mu_i)+\sum_j\delta\mu_jN_j(\mu_i+\delta\mu_i)\ge
\Omega (\mu_i) \label{Eqn4b}\ee

Finally, recalling that $N_i=\partial_{\mu_i}\Omega$ and after
expansion of equation (\ref{Eqn4b}) to second order in
$\delta\mu_i$, we obtain

\be
\frac{\partial^2\Omega}{\partial\mu_i\partial\mu_j}\delta\mu_i\delta\mu_j\le
0, \ee

\noindent hence proving the concavity of the grand-potential (or
conversely the convexity of the pressure).

\end{document}